\documentstyle[a4,11pt]{article}
\begin{document}

\setlength{\baselineskip}{16pt}

\begin{center}

{\bf DISPERSION RELATION FOR ALFV\'{E}N WAVES IN A VISCOUS, DIFFUSIVE 
PLASMA}

\vspace{6mm}

\setlength{\baselineskip}{14pt}

SURESH CHANDRA\footnote{Visiting Associate of IUCAA, Pune, India}

\vspace{1mm}

\noindent
School of Physical Sciences, S.R.T.M. University,\\ Nanded 431 606, India\\
(email: suresh492000@yahoo.co.in)
\end{center}

\setlength{\baselineskip}{16pt}
\noindent
{\bf Abstract.} Propagation of Alfv\'{e}n waves in the solar plasma has been a
topic of scientific interest for a long time. We have derived a dispersion 
relation $\omega^4 + \omega^2 [(\nu^2 + \eta^2) k^4 - v_A^2 k^2] + i \omega 
(\nu + \eta) v_A^2 k^4 + (\nu \eta v_A^2 k^6 + \nu^2 \eta^2 k^8) = 0$ for shear 
Alfv\'{e}n waves in a viscous and diffusive plasma. 

\begin{center}
\rule{5cm}{0.5mm}
\end{center}
                                                                                
The MHD equations for viscous and diffusive plasma are
\begin{eqnarray}
\rho \frac{\partial \hspace{-1mm} \stackrel{\rightarrow}{v}}{\partial t} + \rho 
(\stackrel{\rightarrow}{v}. \nabla)\stackrel{\rightarrow}{v} = \frac{1}{\mu}
(\nabla \times \stackrel{\rightarrow}{B}) \times \stackrel{\rightarrow}{B} + 
\rho \nu \nabla^2 \hspace{-1mm} \stackrel{\rightarrow}{v} \hspace{1cm}
\mbox{Momentum equation} \label{eq:1} \\
\frac{\partial \hspace{-1mm} \stackrel{\rightarrow}{B}}{\partial t} = \nabla 
\times (\stackrel{\rightarrow}{v} \times \stackrel{\rightarrow}{B}) + \eta 
\nabla^2 \hspace{-1mm} \stackrel{\rightarrow}{B} \hspace{3.8cm} 
\mbox{Induction equation} \label{eq:2} \\
\nabla . \stackrel{\rightarrow}{B} = 0 \hspace{5.9cm} \mbox{Magnetic flux 
conservation} \label{eq:3}
\end{eqnarray}

\noindent
where $\stackrel{\rightarrow}{v}$ is the velocity, $\stackrel{\rightarrow}{B}$ 
the magnetic field and $\rho$, $\mu$, $\eta$, $\nu$ are, respectively, the mass 
density, magnetic permeability, magnetic diffusivity and the coefficient of 
viscosity. Let us consider small perturbations from the equilibrium [1]:
\begin{eqnarray}
\rho = \rho_0 + \rho_1 \hspace{2.5cm} \stackrel{\rightarrow}{v} = \stackrel{
\rightarrow}{v}_1 \hspace{2.5cm} \stackrel{\rightarrow}{B} = \stackrel{
\rightarrow }{B}_0 + \stackrel{\rightarrow}{B}_1  \nonumber
\end{eqnarray}

\noindent
and linearize the equations (\ref{eq:1}) through (\ref{eq:3}) by neglecting 
squares and products of the small quantities (denoted by subscript 1). Here, the
quantities with subscript 0 represent their values in equilibrium. Notice that
the equilibrium velocity $v_0$ is zero [1]. After linearization, we 
have the corresponding relations as the following:
\begin{eqnarray}
\rho_0 \frac{\partial \hspace{-1mm} \stackrel{\rightarrow}{v}_1}{\partial t} = 
\frac{1}{\mu} (\nabla \times \stackrel{\rightarrow}{B}_1) \times \stackrel{
\rightarrow}{B}_0 + \rho_0 \nu \nabla^2 \hspace{-1mm} \stackrel{\rightarrow}
{v}_1 \label{eq:4} \\
\frac{\partial \hspace{-1mm} \stackrel{\rightarrow}{B}_1}{\partial t} = \nabla 
\times (\stackrel{\rightarrow}{v}_1 \times \stackrel{\rightarrow}{B}_0) + \eta 
\nabla^2 \hspace{-1mm} \stackrel{\rightarrow}{B}_1 \hspace{0.7cm}\label{eq:5} \\
\nabla . \stackrel{\rightarrow}{B}_1 = 0 \hspace{4.6cm} \label{eq:6}
\end{eqnarray}

\noindent
Here, the magnetic field $\stackrel{\rightarrow}{B}_0$ is taken uniform as well
as time independent. On differentiating equation (\ref{eq:4}) with respect to 
time and then substituting the expressions for $\partial \hspace{-1mm} 
\stackrel{\rightarrow}{v}_1 \hspace{-1mm}/\partial t$ and $\partial 
\hspace{-1mm} \stackrel{ \rightarrow}{B}_1 \hspace{-1mm}/\partial t$ from 
equations (\ref{eq:4}) and (\ref{eq:5}), we get
\begin{eqnarray}
\frac{\partial^2 \hspace{-1mm} \stackrel{\rightarrow}{v}_1}{\partial t^2} = 
v_A^2 \Big\{\nabla \times [\nabla \times (\stackrel{\rightarrow}{v}_1 \times 
\hat{B}_0)] \Big\} \times \hat{B}_0 + \frac{\eta}{\mu \rho_0} \nabla \times 
(\nabla^2 \stackrel{\rightarrow}{B}_1 \times \stackrel{\rightarrow}{B}_0) \hspace{2.0cm} 
\nonumber\\  
+ \frac{\nu}{\mu \rho_0} \nabla^2 [(\nabla \times \stackrel{\rightarrow}{B}_1) 
\times \stackrel{\rightarrow}{B}_0] + \nu^2 \nabla^4 \hspace{-1mm} 
\stackrel{\rightarrow} {v}_1 \label{eq:7}
\end{eqnarray}

\noindent
where $\hat{B}_0$ is the unit vector along $\stackrel{\rightarrow}{B}_0$ and 
$v_A$ ($\equiv B_0/\sqrt{\mu \rho_0})$ is the
Alfv\'{e}n velocity. Let us 
seek a plane-wave solution of the form
\begin{eqnarray}
\stackrel{\rightarrow}{v}_1 = \stackrel{\rightarrow}{v} \ \mbox{e}^{i(\stackrel
{\rightarrow}{k} . \stackrel{\rightarrow}{r} - \omega t)} \hspace{3cm}
\stackrel{\rightarrow}{B}_1 = \stackrel{\rightarrow}{B} \ \mbox{e}^{i(\stackrel
 {\rightarrow}{k} . \stackrel{\rightarrow}{r} - \omega t)} \label{eq:8}
\end{eqnarray}

\noindent
where $\stackrel{\rightarrow}{k}$ is the wave vector and $\omega$ the frequency.
The effect of the plane-wave assumption is simply to replace $\partial/\partial
t$ by $-i \omega$ and $\nabla$ by $i \hspace{-1mm} \stackrel{\rightarrow}{k}$.
Under the plane-wave assumption, equation (\ref{eq:7}) reduces to 
\vspace{-3mm}
\begin{eqnarray}
\omega^2 \stackrel{\rightarrow}{v}_1 = v_A^2 \Big\{\stackrel{\rightarrow}{k}
\times [\stackrel{\rightarrow}{k} \times (\stackrel{\rightarrow}{v}_1 \times 
\hat{B}_0)] \Big\} \times \hat{B}_0 + \frac{i \eta k^2}{\mu \rho_0} 
(\stackrel{\rightarrow}{k} . \stackrel{\rightarrow}{B}_0) \stackrel{\rightarrow}
{B}_1  \hspace{2cm} \nonumber\\
+ \frac{i \nu k^2}{\mu \rho_0} [(\stackrel{\rightarrow}{k} \times \stackrel{
\rightarrow}{B}_1) \times \stackrel{\rightarrow}{B}_0 ] - \nu^2 k^4 \stackrel{
\rightarrow}{v}_1 \label{eq:9}
\end{eqnarray}

\noindent
For plane-wave assumption, equation (\ref{eq:6}) gives
\begin{eqnarray}
\stackrel{\rightarrow}{k} . \stackrel{\rightarrow}{B}_1 = 0 \label{eq:10}
\end{eqnarray}

\noindent
It shows that the magnetic field perturbation is normal to the propagation 
vector $\stackrel{\rightarrow}{k}$. Equation (\ref{eq:9}) can be simplified as
\begin{eqnarray}
\omega^2 \stackrel{\rightarrow}{v}_1 = v_A^2 \Big\{(\stackrel{\rightarrow}{k} .
\hat{B}_0) (\stackrel{\rightarrow}{k} . \hat{B}_0) \stackrel{\rightarrow}{v}_1 -
(\stackrel{\rightarrow}{k} . \hat{B}_0) (\stackrel{\rightarrow}{k} . \stackrel{
\rightarrow}{v}_1) \hat{B}_0 + \stackrel{\rightarrow}{k} \Big[(\stackrel{
\rightarrow}{k} . \stackrel{\rightarrow}{v}_1) - (\hat{B}_0 . \stackrel{
\rightarrow}{v}_1) (\stackrel{\rightarrow}{k} . \hat{B}_0) \Big]\Big\} \nonumber
\end{eqnarray}
\vspace*{-6mm}
\begin{eqnarray} 
+ \frac{i \eta k^2}{\mu \rho_0} (\stackrel{\rightarrow}{k} . \stackrel{
\rightarrow}{B}_0) \stackrel{\rightarrow}{B}_1 + \frac{i \nu k^2}{\mu \rho_0} 
\Big[(\stackrel{\rightarrow}{B}_1 ( \stackrel{ \rightarrow}{k} . \stackrel{ 
\rightarrow}{B}_0) - \stackrel{\rightarrow}{k} (\stackrel{ \rightarrow}{B}_0 . 
\stackrel{ \rightarrow}{B}_1)\Big] - \nu^2 k^4 \stackrel{ \rightarrow}{v}_1 
\label{eq:11}
\end{eqnarray}

\noindent
Let the propagation vector  $\stackrel{ \rightarrow}{k}$ makes an angle $\theta$
with the equilibrium magnetic field $\stackrel{ \rightarrow}{B}_0$. Then 
equation (\ref{eq:11}) reduces to
\begin{eqnarray}
\omega^2 \stackrel{\rightarrow}{v}_1 = v_A^2 \Big\{k^2 \mbox{cos}^2 \theta 
\stackrel{ \rightarrow}{v}_1 - k \ \mbox{cos} \ \theta (\stackrel{\rightarrow}
{k} . \stackrel{\rightarrow}{v}_1) \hat{B}_0  + \stackrel{\rightarrow}{k} \Big[
(\stackrel{\rightarrow}{k} . \stackrel{\rightarrow}{v}_1) - k \ \mbox{cos} \ 
\theta (\hat{B}_0 . \stackrel{\rightarrow}{v}_1)\Big]\Big\} \nonumber 
\end{eqnarray}
\vspace*{-6mm}
\begin{eqnarray}
+ \frac{i \eta k^3 B_0 \mbox{cos} \ \theta \ \stackrel{\rightarrow}{B}_1}
{\mu \rho_0} + \frac{i \nu k^2}{\mu \rho_0} \Big[k \ B_0 \ \mbox{cos} \ 
\theta \stackrel{\rightarrow}{B}_1 - \stackrel{\rightarrow}{k} (\stackrel{ 
\rightarrow}{B}_0 . \stackrel{\rightarrow}{B}_1)\Big] - \nu^2 k^4 \stackrel{
\rightarrow}{v}_1 \label{eq:12}
\end{eqnarray}

\noindent
Multiplying equation (\ref{eq:12}) by $\hat{B}_0$, we get
\begin{eqnarray}
 \omega^2 (\stackrel{\rightarrow}{v}_1 . \hat{B}_0) = \frac{i \eta k^3 \cos 
\theta}{\mu \rho_0} (\stackrel{\rightarrow}{B}_1 . \stackrel{\rightarrow}{B}_0)
- \nu^2 k^4 (\stackrel{\rightarrow}{v}_1 . \hat{B}_0)    \nonumber     
\end{eqnarray}
                                                                       
\noindent
The imaginary part disappears when we consider $\stackrel{\rightarrow}{B}_1$
orthogonal to $\stackrel{\rightarrow}{B}_0$. Now, we get
\begin{eqnarray}
(\omega^2 + \nu^2 k^4) (\hat{B}_0 . \stackrel{\rightarrow}{v}_1) = 0 
 \label{eq:13}
\end{eqnarray}

\noindent
Since $(\omega^2 + \nu^2 k^4)$ is not zero, we have 
\vspace{-3mm}
\begin{eqnarray}
\hat{B}_0 . \stackrel{\rightarrow}{v}_1 = 0 \label{eq:14}
\end{eqnarray}

\noindent
It shows that the perturbation of velocity is normal to the ambient magnetic 
field. 
The discussion so far leads to the geometry as shown in Figure \ref{geometry1}.

\newpage
                                                                                
\vspace*{33mm}
                                                                                
\hspace{6cm}
\begin{picture}(0,0)
\bezier{120}(0,0)(40,0)(80,0)
\put(83,-5){$\stackrel{\rightarrow}{B}_1$}
                                                                                
\bezier{120}(0,0)(0,40)(0,80)
\put(-5,83){$\stackrel{\rightarrow}{v}_1$}
                                                                                
\bezier{50}(0,0)(-15,-15)(-30,-30)
\put(-40,-40){$\stackrel{\rightarrow}{B}_0$}
                                                                                
\bezier{45}(0,0)(-15,20)(-30,40)
\put(-40,45){$\stackrel{\rightarrow}{k}$}
                                                                                
\bezier{30}(-10,-10)(-15,0)(-10,10)
\put(-20,0){$\theta$}
\end{picture}
                                                                                
\vspace{1.0cm}
                                                                                
\begin{figure}[h]
\caption{\small $\stackrel{\rightarrow}{B}_0$, $\stackrel{\rightarrow}{B}_1$
and $\stackrel{\rightarrow}{v}_1$ are orthogonal to each other. $\stackrel{
\rightarrow}{k}$ lies in the ($\stackrel{\rightarrow}{B}_0$, $\stackrel{
\rightarrow}{v}_1$)-plane and $\stackrel{\rightarrow}{k}$ makes and angle
$\theta$ with $\stackrel{\rightarrow}{B}_0$.}\label{geometry1}
\end{figure}

Now, multiplying equation (\ref{eq:12}) by $\stackrel{\rightarrow}{k}$
and using equations (\ref{eq:10}) and (\ref{eq:14}), we get
\begin{eqnarray}
\omega^2 (\stackrel{\rightarrow}{v}_1 . \stackrel{\rightarrow}{k}) = v_A^2 k^2 
(\stackrel{\rightarrow}{v}_1 . \stackrel{\rightarrow}{k}) - \nu^2 k^4 
(\stackrel{\rightarrow}{v}_1 . \stackrel{\rightarrow}{k}) \nonumber \\
\nonumber \\
\Big[\omega^2 - v_A^2 k^2 + \nu^2 k^4\Big](\stackrel{\rightarrow}{k} . 
\stackrel{\rightarrow}{v}_1) = 0 \hspace{1.7cm} \label{eq:18}
\end{eqnarray}

Thus, we have either
\begin{eqnarray}
\stackrel{\rightarrow}{k} . \stackrel{\rightarrow}{v}_1 = 0 \label{eq:19}
\end{eqnarray}
                                                                                
\noindent
or
\begin{eqnarray}
\omega^2 - v_A^2 k^2 + \nu^2 k^4 = 0 \label{eq:20}
\end{eqnarray}
                                                                                
\noindent
The relation (\ref{eq:19}) leads to the shear Alfv\'{e}n waves whereas the
relation (\ref{eq:20}) leads to the compressional Alfv\'{e}n waves.
For shear Alfv\'{e}n waves, the perturbation is incompressible so that we have
$\nabla . \stackrel{\rightarrow}{v}_1 = 0$. Under the plane-wave assumption,
this relation gives the equation (\ref{eq:19}).

Since $\stackrel{\rightarrow}{k} . \stackrel{\rightarrow}{v}_1 = 0$, we have 
$\theta = 0$. Using equation (\ref{eq:19}) (with $\theta = 0$) in
(\ref{eq:12}) (for Figure \ref{geometry1}), we get 
\begin{eqnarray}
\omega^2 \stackrel{\rightarrow}{v}_1 = v_A^2 k^2 \stackrel{\rightarrow}{v}_1 
+ \frac{i (\nu + \eta) k^3 B_0}{\mu \rho_0} \stackrel{\rightarrow}{B}_1 -
\nu^2 k^4 \stackrel{\rightarrow}{v}_1\label{eq:25}
\end{eqnarray}

For the plane wave solution, equation (\ref{eq:5}) gives
\begin{eqnarray}
(\eta k^2 - i \omega) \stackrel{\rightarrow}{B}_1 = i B_0 k \stackrel{
\rightarrow}{v}_1  \label{eq:28}
\end{eqnarray}
                                                                                
\noindent
From equations (\ref{eq:25}) and (\ref{eq:28}), we get a dispersion
relation                                                                       
\begin{eqnarray}
\omega^4 + \omega^2 [(\nu^2 + \eta^2) k^4 - v_A^2 k^2] + i \omega (\nu + \eta) 
v_A^2 k^4 + (\nu \eta v_A^2 k^6 + \nu^2 \eta^2 k^8) = 0  \label{eq:29} 
\end{eqnarray}

\noindent
This dispersion relation incorporates the viscosity as well as diffusivity.

For the plane wave solution, equation (\ref{eq:4}) gives
\begin{eqnarray}
(\nu k^2 - i \omega) \stackrel{\rightarrow}{v}_1 = \frac{i B_0 k}{\mu \rho_0} \
\stackrel{\rightarrow}{B}_1  \label{eq:26}
\end{eqnarray}

\noindent
From equations (\ref{eq:25}) and (\ref{eq:26}), we get a dispersion relation
\begin{eqnarray}
\omega^2 = k^2 \Big[v_A^2  - i \omega (\nu + \eta)\Big] + \nu \eta k^4
\nonumber     
\end{eqnarray}

\noindent
This dispersion relation is the same as obtained by Pek\"unl\"u et al. [2] 
and can also be obtained by using the equations (\ref{eq:26}) and (\ref{eq:28}). 

We have thus obtained a new dispersion relation (\ref{eq:29}).

\begin{center}
{\bf Acknowledgments}
\end{center}
 We heartily thank Prof. E.R. Pek\"unl\"u for 
cooperative and encouraging correspondence. Financial support from the 
Department of Science \& Technology, New Delhi and the Indian Space Research 
Organization (ISRO), Bangalore in the form of research projects is thankfully
acknowledged.

\begin{center}
{\bf References}
\end{center}
\begin{description}
\item{} [1] E.R. Priest, {\it Solar Magnetohydrodynamics} (1982) D. Reidel
Publishing Company, Dordrecht, Holland.

\item{} [2] E.R. Pek\"unl\"u, Z. Bozkurt, M. Afsar, E. Soydugan and F. Soydugan,
{\it Mon. Notices Roy. Astron. Soc.} {\bf 336} (2002) 1195.
\end{description}

\end{document}